# Long COVID Prevalence, Disability, and Accommodations: Analysis Across Demographic Groups

*Jennifer Cohen, Miami University, Oxford, OH, USA and*
*University of the Witwatersrand, South Africa*
*ORCID 0000-0001-7131-8372*
*Yana Rodgers, Rutgers University, New Brunswick, NJ, USA**
*ORCID 0000-0001-7669-2857*

**Abstract**

**Purpose:** This paper examines the prevalence of long COVID across different demographic groups in the U.S. and the extent to which workers with impairments associated with long COVID have engaged in pandemic-related remote work.

**Methods:** We use the U.S. Household Pulse Survey to evaluate the proportion of all adults who self-reported to (1) have had long COVID, and (2) have activity limitations due to long COVID. We also use data from the U.S. Current Population Survey to estimate linear probability regressions for the likelihood of pandemic-related remote work among workers with and without disabilities.

**Results:** Findings indicate that women, Hispanic people, sexual and gender minorities, individuals without four-year college degrees, and people with preexisting disabilities are more likely to have long COVID and to have activity limitations from long COVID. Remote work is a reasonable arrangement for people with such activity limitations and may be an unintentional accommodation for some people who have undisclosed disabilities. However, this study shows that people with disabilities were less likely than people without disabilities to perform pandemic-related remote work.

**Conclusion:** The data suggest this disparity persists because people with disabilities are clustered in jobs that are not amenable to remote work. Employers need to consider other accommodations, especially shorter workdays and flexible scheduling, to hire and retain employees who are struggling with the impacts of long COVID.

**Acknowledgments:** Author YR has received funding support from the National Institute on Disability, Independent Living, and Rehabilitation Research (NIDILRR) for the Rehabilitation Research & Training Center (RRTC) on Employment Policy: Center for Disability-Inclusive Employment Policy Research Grant [grant number #90RTEM0006-01–00] and by the NIDILRR RRTC on Employer Practices Leading to Successful Employment Outcomes Among People with Disabilities Research Grant [grant number #90RTEM0008-01-00].

* Corresponding Author Yana Rodgers (yana.rodgers@rutgers.edu).

## I. Introduction

Early in the pandemic, reports emerged of patients with persistent symptoms after being infected with COVID-19. Members of this patient community called themselves "Long Haulers" and referred to their prolonged illness as "long COVID" [1]. The U.S. Centers for Disease Control subsequently defined long COVID formally as "a range of new, returning, or ongoing health problems lasting four or more weeks after first being infected with COVID-19" [2]. Symptoms include difficulty thinking, remembering, or concentrating (often referred to as brain fog), depression, anxiety, headaches, shortness of breath, heart palpitations, and fatigue. These symptoms can be mild at best and debilitating at worst. Notably, long COVID is not a single condition. Rather, it entails multiple, potentially simultaneous pathologies that may have different causes and risk factors and could manifest as new chronic diseases such as heart disease, diabetes, and mental and neurological conditions that prevent patients from returning to their previous state of health [1]. Together with the U.S. Department of Justice, the U.S. Department of Health and Human Services (DHHS) declared that long COVID can qualify as a disability under the Americans with Disabilities Act and thus be eligible for reasonable accommodation[1] [3].

Estimates published in August 2022 cited in official U.S. government communication indicate that anywhere from 5% to 30% of people who were infected with COVID-19 developed long COVID, amounting to between 7.7 and 23 million individuals in the U.S. [4,5]. Moreover, COVID-related disabilities already far outnumber COVID deaths, and the differential is expected to grow over time [6]. Long COVID could cost the U.S. economy $997 billion in reduced

[1] According to this DHHS guidance, a person with a disability is defined as someone having a physical or mental impairment that substantially limits at least one of their major life activities; and a reasonable accommodation is an adjustment of an organization's policies, practices, or workspace that allows people with disabilities to have equal access to opportunities and employment.

earnings from people working less and over $2 trillion in reduced quality of life [7]. Data for a large global sample covering 56 countries indicate that the vast majority (85%) of respondents with long COVID experienced brain fog (cognitive dysfunction or memory issues), followed closely by fatigue and post-exertional malaise [8]. Almost half of these respondents needed accommodations for shorter workdays, and about one quarter were not working due to their illness – they had either taken a leave, were dismissed by their employer, or could not find a job with the appropriate accommodations [8].

These aggregate statistics, however, hide large disparities between demographic groups. Emerging research indicates that social determinants of health appear to play a large role in COVID-19 exposure and infection rates [9], but less is known about long COVID. To fill this knowledge gap, this paper examines the prevalence of long COVID and activity limitations due to long COVID across demographic groups, with a focus on gender identity, race, ethnicity, sexual orientation, age, education, and preexisting disabilities. We also explore appropriate accommodations in the workplace for people with such activity limitations. The analysis pays particular attention to how occupational segregation affects who contracts COVID to begin with and who is able to work remotely. We consider how remote work may be an *unintentional accommodation* for some white-collar workers with undisclosed disabilities. However, because remote work is often not an option for blue collar workers and many in the service sector – areas where people with disabilities are overrepresented – we conclude with a discussion of workplace accommodations aside from remote work that are particularly relevant for people with activity limitations arising from long COVID.

## II. Data and Methodology

The first part of our study uses the U.S. Household Pulse Survey (HPS) to evaluate the proportion of all adults ages 18 and above who self-reported to (1) have had long COVID, and (2) have activity limitations due to long COVID. The HPS is a rapid deployment, rapid dissemination bi-weekly survey about COVID and emergent issues to inform federal and state government response and recovery, covering between 40,000 and 75,000 individuals in each wave. In June 2022, the Census Bureau added questions about long COVID to the HPS, giving researchers a better understanding of the condition's prevalence. The two survey questions about long COVID that we used for our study are: "Did you have any symptoms lasting 3 months or longer that you did not have prior to having coronavirus or COVID-19?" with a yes/no answer; and "Do these long-term symptoms reduce your ability to carry out day-to-day activities compared with the time before you had COVID-19?" with the responses coded along a 3-point scale of "Yes, a lot; Yes, a little; Not at all," thus capturing information about the severity of activity limitations as well.[2] The survey also collects demographic data on gender identity, race, ethnicity, sexuality, level of education, and disability status.[3] We report average long COVID prevalence for 2022 (7 survey rounds between June 2022 and December 2022), and we report average activity limitations for all periods in which the question was asked (7 rounds between September 2022 and March 2023). The average weighted response rate for the survey rounds we used is 5.7%.[4] Data on long COVID prevalence are reported as overall means (for all adults) and as conditional means (for adults who ever had a COVID infection).

The second part of our study uses the U.S. Current Population Survey (CPS) to examine the likelihood of remote work. Although the HPS has a question about remote work, the data on long COVID and remote work only go back to June 2022. In contrast, the CPS has monthly data on pandemic-related remote work dating back to May 2020 and extending through September 2022. We also prefer to use the CPS to examine the likelihood of remote work because the CPS has far more detailed information on individuals' occupation and industry of employment, while the HPS includes only broad industry categories and no information at all on occupation. We refer to the remote work variable in the CPS as pandemic-related remote work because the survey asked whether the respondent did paid work from home because of the pandemic at any time in the past 4 weeks. Hence this measure specifically captures whether the COVID-19 pandemic caused the respondent to work remotely for any reason. We do recognize that this measure has its limitations, including the fact that it does not capture whether or not the respondent normally works remotely, nor does it capture current trends in hybrid working that may be better measured with a Likert scale [10].

[2] https://www.cdc.gov/nchs/covid19/pulse/long-covid.htm
[3] The data and questionnaires are publicly available at: https://www.census.gov/programs-surveys/household-pulse-survey/datasets.html.
[4] Computed from response rates published at https://www.cdc.gov/nchs/covid19/pulse/long-covid.htm.

Although the CPS does not have information on long COVID per se, it does include measures of disability based on a six-question set: (1) "Is this person deaf or does he/she have serious difficulty hearing?"; (2) "Is this person blind or does he/she have serious difficulty seeing even when wearing glasses?"; (3) "Because of a physical, mental, or emotional condition, does this person have serious difficulty concentrating, remembering, or making decisions?"; (4) "Does this person have serious difficulty walking or climbing stairs?"; (5) "Does this person have difficulty dressing or bathing?"; (6) "Because of a physical, mental, or emotional condition, does this person have difficulty doing errands alone such as visiting a doctor's office or shopping?" Respondents may respond to more than one question in the affirmative, so the corresponding categories of disability are not mutually exclusive.

Using the CPS, we run linear probability regressions to predict the likelihood of pandemic-related remote work among all workers. Every model includes a variable for survey month/year to capture the trend over time. We perform the first regression with disability status only, where the disability variable indicates having at least one disability. We then estimate a second model that includes disability status plus occupations (524 categories) and industries (51 categories). The second model shows the extent to which the jobs that people hold (occupations) and the sectors in which they work (industries) can explain whether or not someone is working remotely. Our third model includes disability status plus demographic characteristics (gender, age, race/ethnicity, education, number of children under age 18 in the household, part-time versus full-time status, and employee versus self-employed status), without the occupation and industry controls. The fourth model includes disability status plus the demographic characteristics together with the detailed occupations and industries. Finally, we estimate a fifth regression using the six separate measures of disability instead of the overall disability indicator, plus the complete set of control variables.

## III. Prevalence and Repercussions of Long COVID

Our analysis of the HPS data on long COVID indicates that 14.3% of all adults in the U.S. have had long COVID (Figure 1, Panel A), and conditional on having contracted COVID, 31.1% of adults developed long COVID (Figure 1, Panel B). Both Panels A and B show that women, younger people, Hispanic individuals, sexual and gender minorities (SGM)[5], those without a 4-year degree, and people with at least one preexisting disability[6] are more likely than their counterparts to experience long COVID. Among people who had COVID (Panel B), 50.4% of people with a preexisting disability and 48.2% of transgender people developed long COVID. Focusing on gender, 17.4% of all women had long COVID compared with 10.8% of all men, and conditional on having had a COVID infection, 36.3% of women and 24.5% of men developed long COVID. Our results for the disproportionate representation of women among adults with long COVID are consistent with findings in other studies based on health surveys [12, 13]. Men and college graduates were the least likely to develop long COVID, although a nontrivial number of people from both groups still did.

[5] People who are SGM self-identified their sexual orientation as gay, bisexual, "something else," or gender identification as transgender in the HPS survey [11]. The combined SGM category is heterogeneous; we do not mean to imply that gay, bisexual, and transgender people have comparable experiences in daily life or in the labor market.

[6] The disability data in the HPS may include people with disabilities from long COVID, but the majority of disabilities were likely preexisting disabilities, as indicated by the trend analysis in Section 3.

**Figure 1.** Prevalence of Long Covid by Demographic Groups, 2022 Average

Panel A. Percentage of All Adults Who Ever Had Long COVID

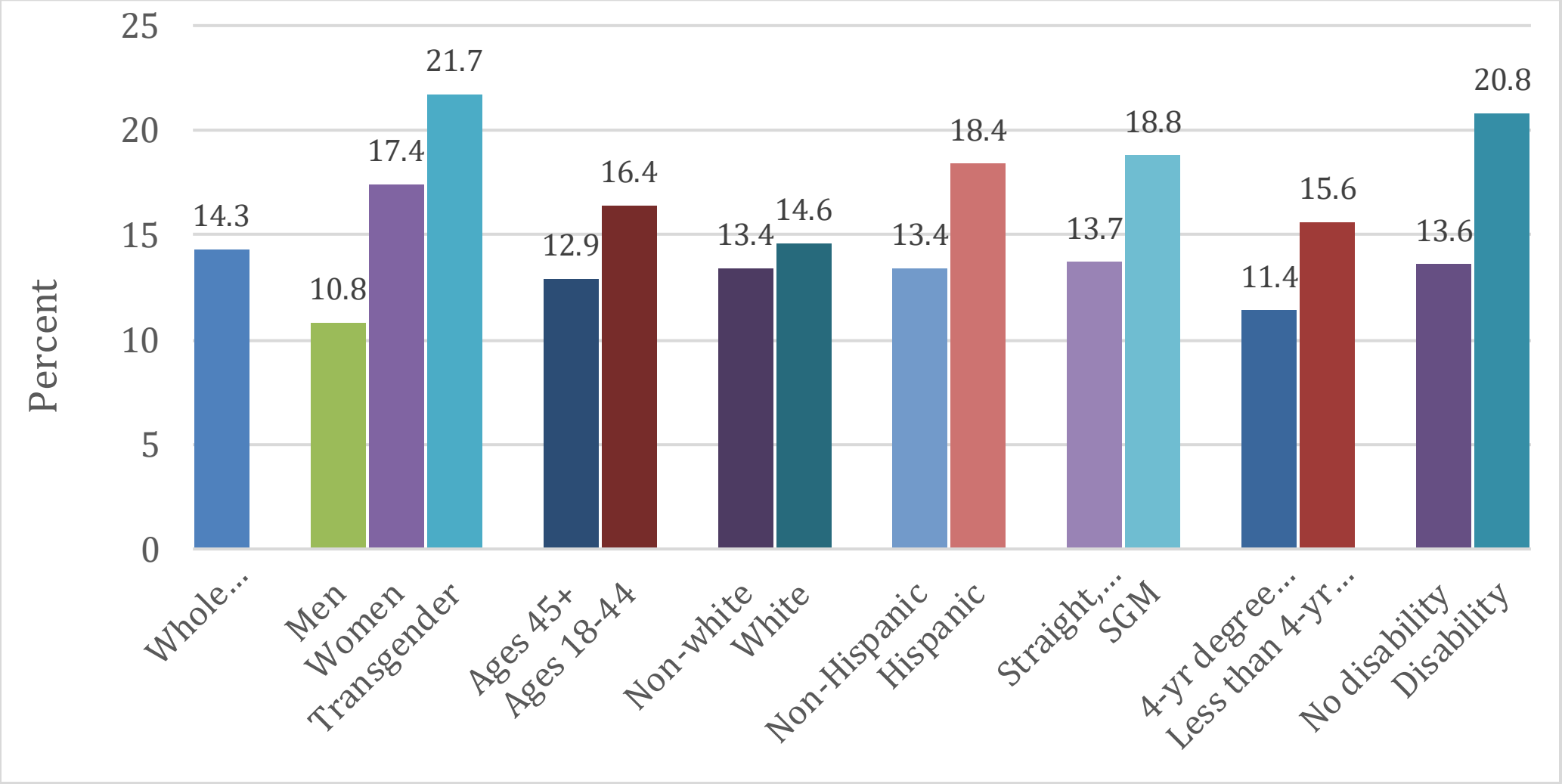

Panel B. Percentage of Adults who had COVID Who Developed Long COVID

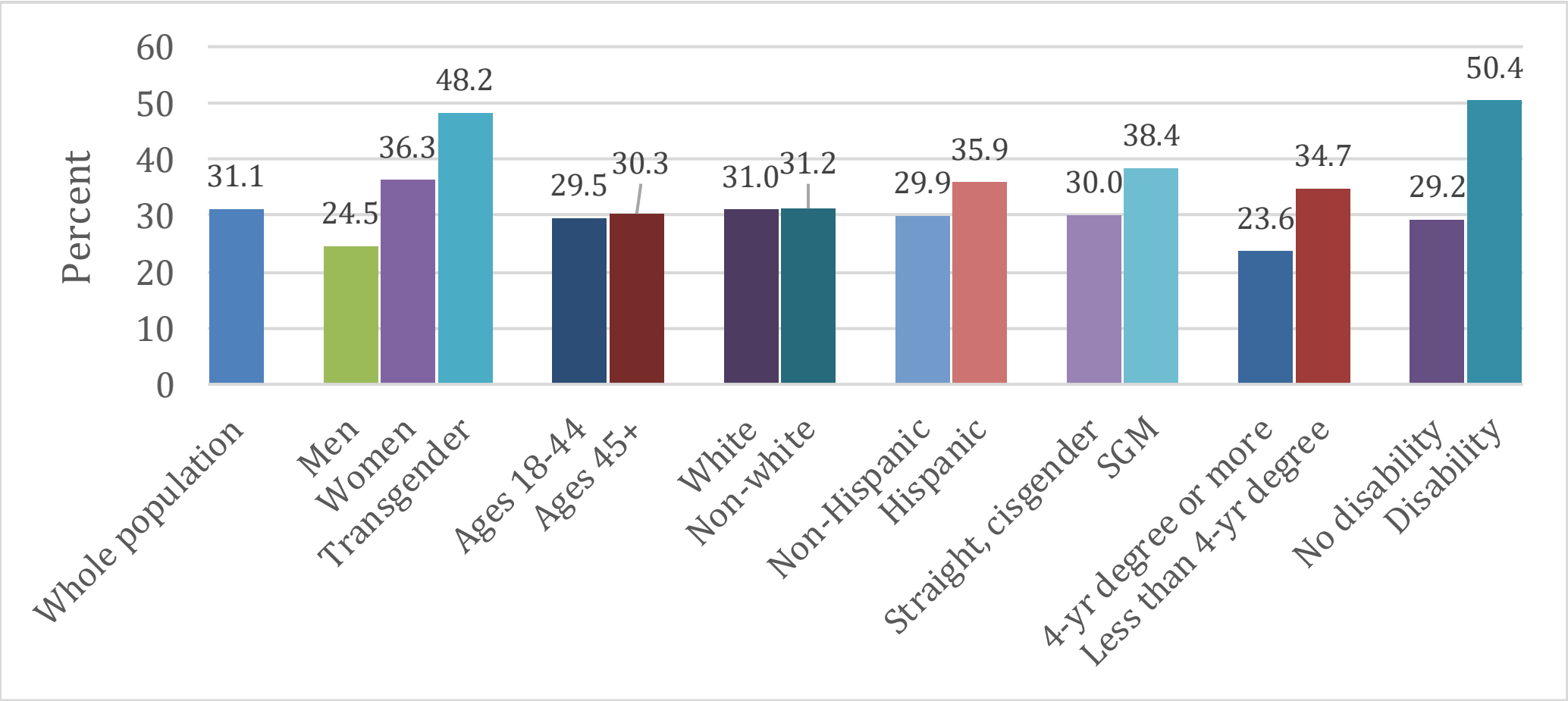

Constructed by the authors using HPS data for adults ages 18 and above, June-Dec. 2022. All group means are statistically significantly different from each other at 5% level or better, except Panel B for Ages 18-44/Ages 45+ and White/Non-white. T-statistics and probability levels in parentheses for mean differences across Panel A are 26.02 (0.000) for women, 3.44 (0.001) for transgender, 14.81 (0.000) for ages 18-44, 4.17 (0.000) for White, 11.39 (0.000) for Hispanic, 11.27 (0.000) for SGM, 21.20 (0.000) for less than 4-year degree, and 15.16 (0.000) for disability. T-statistics and probability levels in parentheses for mean differences across Panel B are 23.40 (0.000) for women, 4.81 (0.000) for transgender, -1.48 (0.140) for ages 18-44, -0.27 (0.787) for White, 7.80 (0.000) for Hispanic, 10.60 (0.000) for SGM, 28.25 (0.000) for less than 4-year degree, and 23.94 (0.000) for disability.

Long COVID can have persistent physical, mental, emotional, and neurological repercussions for one's ability to participate in day-to-day activities. Just as long COVID prevalence is stratified across demographic groups, activity limitations in daily life from long COVID are also distributed along the lines of gender, disability status, education, and SGM status (Figure 2). Women were more likely than men to have activity limitations due to long COVID (11.8% versus 7.3%), and the same is true for less educated individuals compared to those with at least a four-year degree (11.1% versus 7.6%) and SGM compared to straight, cisgender people (14.1% versus 9.3%). The groups most impacted are people with a disability (23.6%) and people who are transgender (24.2%). The same groups reported severe activity limitations (Figure 3).

**Figure 2.** Prevalence of Any Activity Limitations Due to Long COVID by Demographic Group

Panel A. Percentage of All Adults

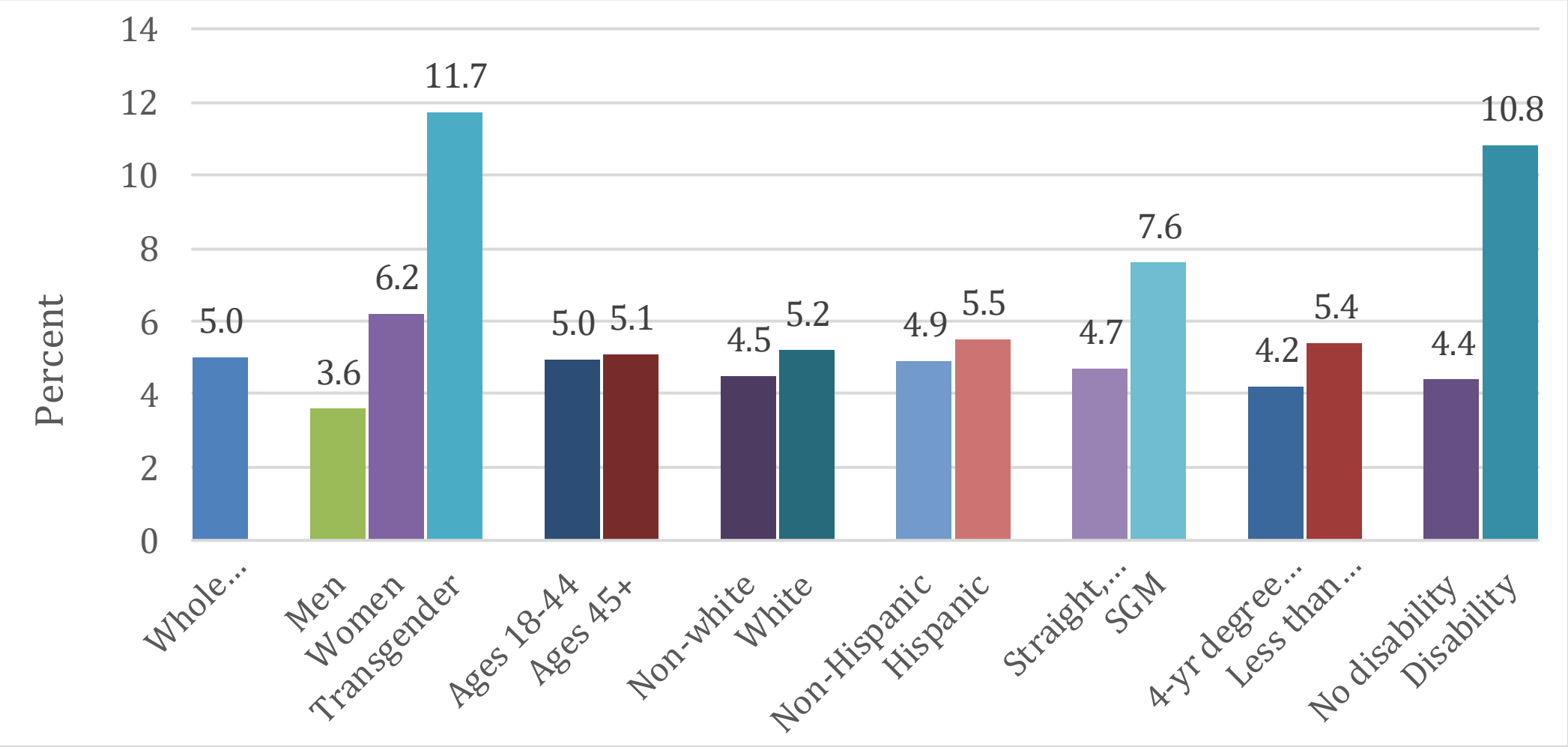

Panel B. Percentage of Adults who had COVID

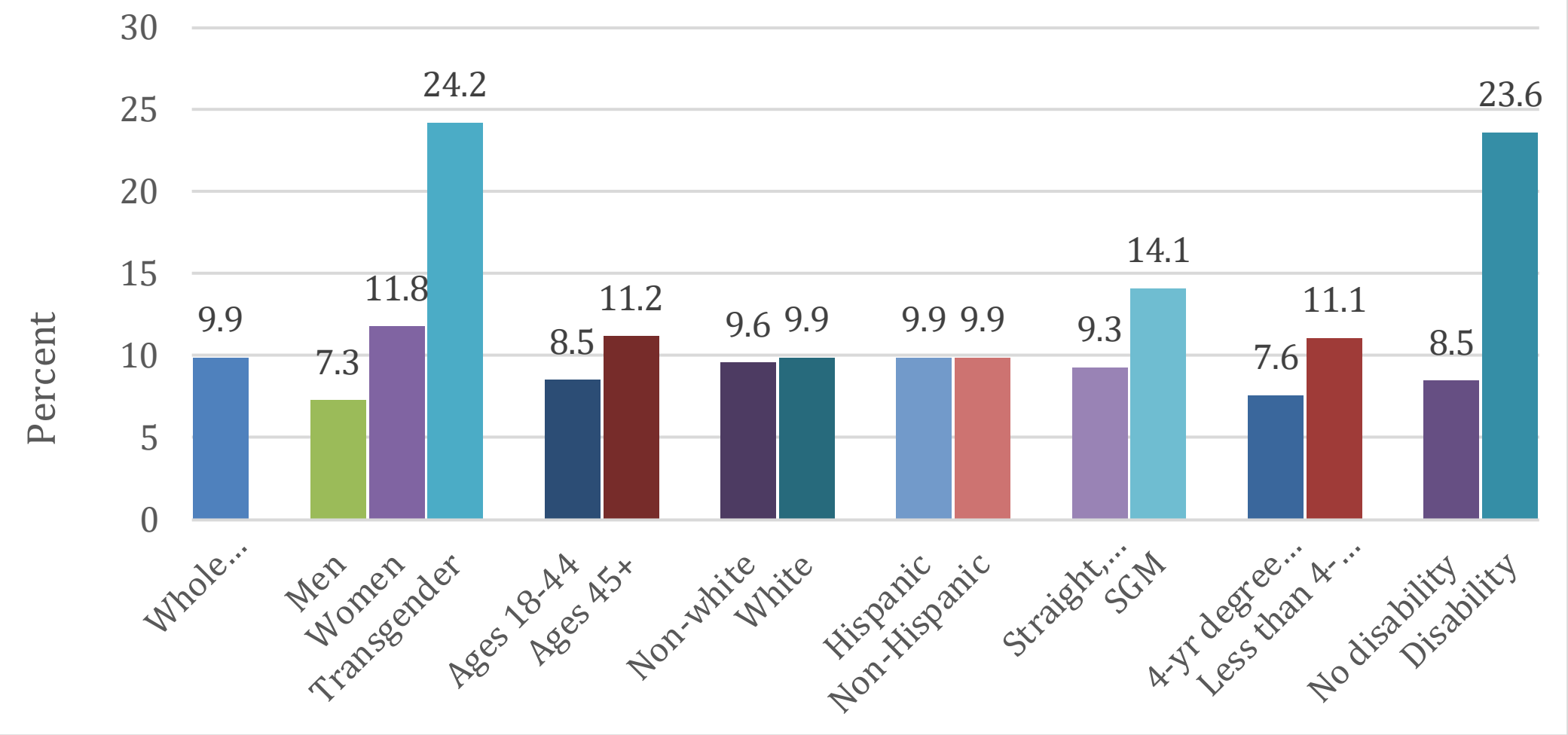

Constructed by the authors using HPS data for adults ages 18 and above, Sep. 2022-Mar. 2023. All group means are statistically significantly different from each other at 5% level or better, except Panel A for Ages 18-44/Ages 45+, and Panel B for White/Non-white and Hispanic/Non-Hispanic. T-statistics and probability levels in parentheses for mean differences across Panel A are 17.78 (0.000) for women, 4.37 (0.000) for transgender, -1.24 (0.217) for ages 18-44, 4.76 (0.000) for White, 2.19 (0.029) for Hispanic, 10.23 (0.000) for SGM, 10.43 (0.000) for less than 4-year degree, and 19.33 (0.000) for disability. T-statistics and probability levels in parentheses for mean differences across Panel B are 15.39 (0.000) for women, 5.13 (0.000) for transgender, -10.60 (0.000) for ages 18-44, 1.01 (0.314) for White, 0.08 (0.934) for Hispanic, 9.52 (0.000) for SGM, 15.51 (0.000) for less than 4-year degree, and 22.68 (0.000) for disability.

Because the HPS data on activity limitations are self-reported, our estimates could be biased in either direction due to self-report bias [14]. As an example of why our estimates may be underestimations, some people experience long COVID effects that do not impact daily activities but nonetheless pose serious challenges. For example, a person who did not exercise on a daily basis may find that they are persistently short of breath when they do try to exercise after long COVID, with implications for co-morbidities like diabetes, anxiety, and depression. Others who do not consider shopping a daily activity may find that they are short of breath when they go shopping at their local mall and decide to curtail their excursions. Brain fog may come and go; or people may go to great lengths to keep it from interfering in activities. In this case, the HPS estimates could understate actual activity limitations because people may be less likely to report limitations on less usual activities and constraints that preclude taking up new health-oriented habits.

**Figure 3.** Prevalence of Severe Activity Limitations Due to Long COVID by Demographic Group

Panel A. Percentage of All Adults

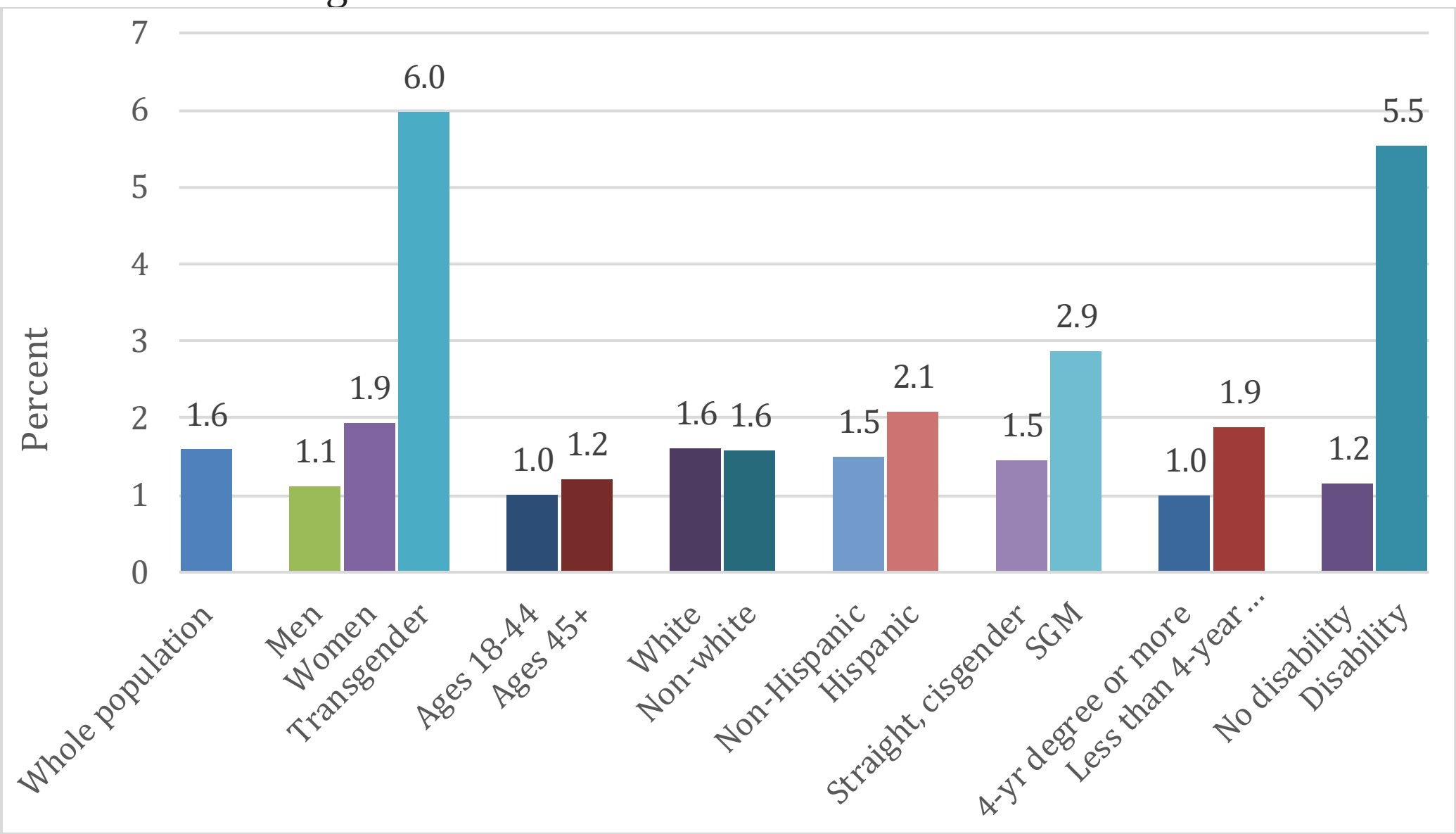

Panel B. Percentage of Adults who had COVID

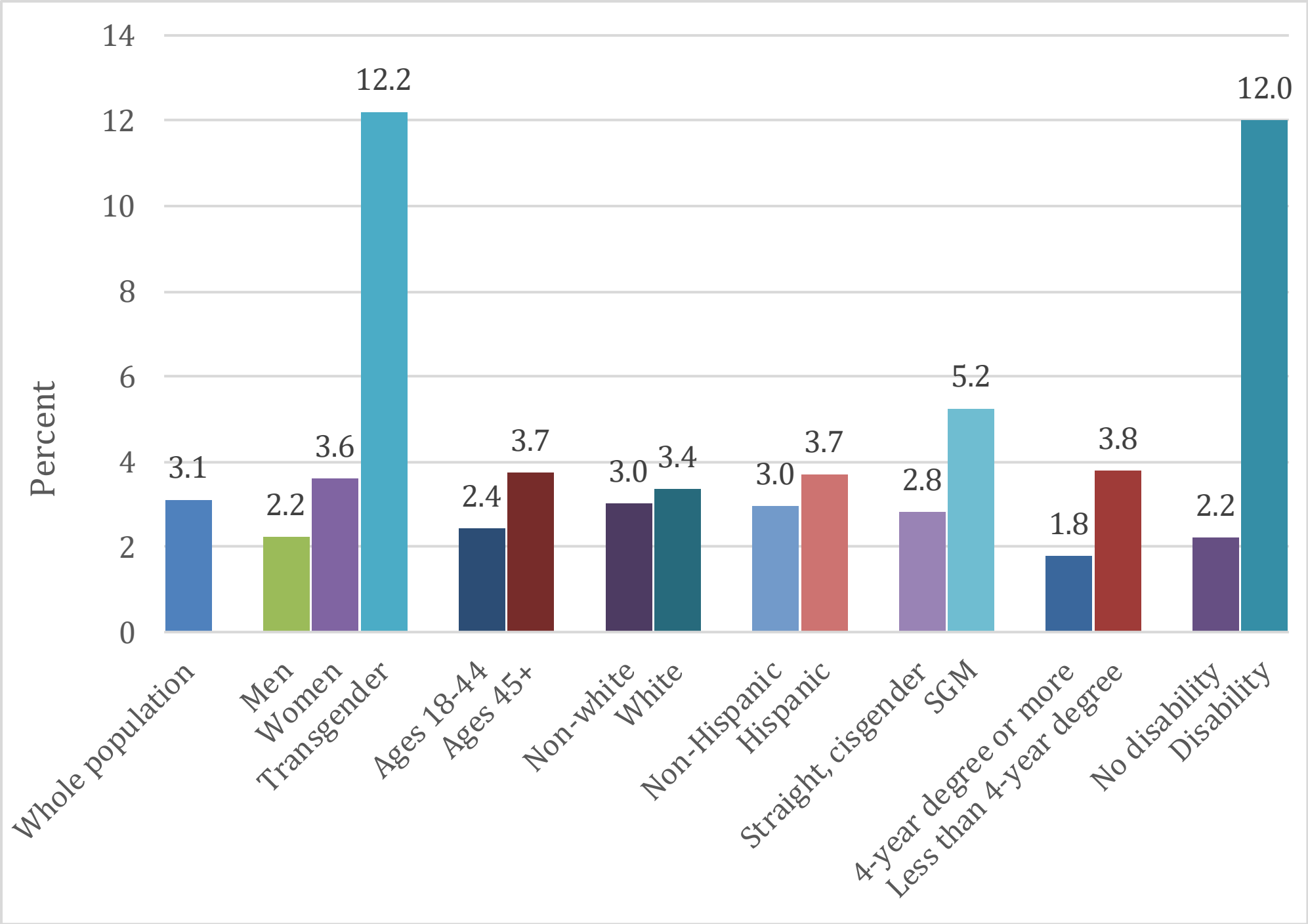

Constructed by the authors using HPS data for adults ages 18 and above, Sep. 2022-Mar. 2023. All group means are statistically significantly different from each other at 5% level or better, except Panels A and B White/Non-white. T-statistics and probability levels in parentheses for mean differences across Panel A are 7.74 (0.000) for women, 3.41 (0.001) for transgender, -3.36 (0.001) for ages 18-44, 0.18 (0.861) for White, 3.33 (0.001) for Hispanic, 6.89 (0.000) for SGM, 12.75 (0.000) for less than 4-year degree, and 16.47 (0.000) for disability. T-statistics and probability levels in parentheses for mean differences across Panel B are 6.42 (0.000) for women, 3.69 (0.000) for transgender, -

7.96 (0.000) for ages 18-44, -1.72 (0.085) for White, 2.47 (0.014) for Hispanic, 6.55 (0.000) for SGM, 14.87 (0.000) for less than 4-year degree, and 17.78 (0.000) for disability.

Long COVID prevalence also varies by the industries in which people are employed. The data on broad industry categories in the HPS indicate that over one third (36.1%) of people in the social assistance industry developed long COVID, and in-industry prevalence was especially high among women (41.6%) relative to men (30.5%). By contrast, 22% of workers in Information Technology (IT) – an industry that is amenable to remote work – developed long COVID. In addition to IT, construction and manufacturing are also male-dominated, meaning that men were better able to protect themselves because of occupational segregation; some by working from home and others by the nature of the work itself. Only 17.4% of men in finance and insurance, another industry that is amenable to remote work, developed long COVID. The ability to work from home and occupational segregation appear to have protected men more than women, as long COVID is more prevalent among women in IT (28.1%), finance and insurance (29.8%), manufacturing (33.5%), and construction (37.1%) than it is among men. These prevalence rates suggest the presence of occupational segregation within industries as well. The HPS does not contain information on specific jobs, but one industry – accommodation and food services – demonstrates the impacts of gendered occupational segregation: 36.9% of women and 20.8% of men in the industry developed long COVID. In this industry, women are more likely to work customer-facing jobs that put them in contact with more people and pathogens, including jobs such as grocery store clerks, fast-food restaurant cashiers, cafeteria workers, hotel desk clerks, and housekeepers [15].

The distribution of work by gender is also uneven within the home, which can help to explain the higher prevalence of long COVID among women, even in industries with lower prevalence rates. Because women perform a disproportionate amount of care work within the home, even during the pandemic, it is likely that relatively more women were exposed in their own homes and then went on to develop long COVID. Notably, of the HPS respondents who were not in paid employment, women (88.8%) were almost 10 times as likely as men (9.3%) to identify caring for children who were not in school or daycare as the primary reason for not being employed.

## IV. Likelihood of Remote Work

Thus far our results indicate that women, sexual and gender minorities, people without a four-year college degree, and people with preexisting disabilities can disproportionately benefit from disability-antidiscrimination legislation and Americans with Disabilities Act (ADA)-mandated accommodations for workers with long COVID [16]. Remote work, often called working from home, is a major accommodation that we consider in our analysis. Earlier research indicates that remote work can especially benefit individuals with cognitive/mental health issues who may value being away from a stressful environment and need to take unscheduled breaks [17]. Another key benefit of remote work is flexibility, which is of particular value for people who have mental and mobility impairments from long COVID that make it more challenging to work in traditional workplace settings. Therefore we are interested in estimating the likelihood of remote work during the pandemic and how that differs between people with and without disabilities.

For this part of the analysis we shift to using the CPS because it has much richer information on respondents' occupations and industries of employment. Although the CPS does not have a direct question on long COVID, we do know about different types of physical and mental disabilities. Calculations using the CPS show a large increase in the percentage of U.S. adults with at least one disability as the pandemic progressed, from a low of 11.3% in July 2020 to as high as 12.8% in September 2022 (Figure 4). According to Sheiner and Salwati (2022), the increase in the percentage of adults with disabilities during the pandemic relative to the 2017-2019 pre-pandemic trend offers an approximation of the extent to which the subsequent increase captures disability related to long COVID [18]. Using this proxy, the authors estimate that there are approximately 2.1 million more people with disabilities attributed to long COVID. We applied the Sheiner and Salwati (2022) method to CPS data for January 2017 to January 2023, keeping in mind that this proxy focuses on people who have serious difficulties with particular tasks and is not a direct estimate of total long COVID prevalence.

**Figure 4.** Percent of U.S. Adults with Any Disability, 2017-2023

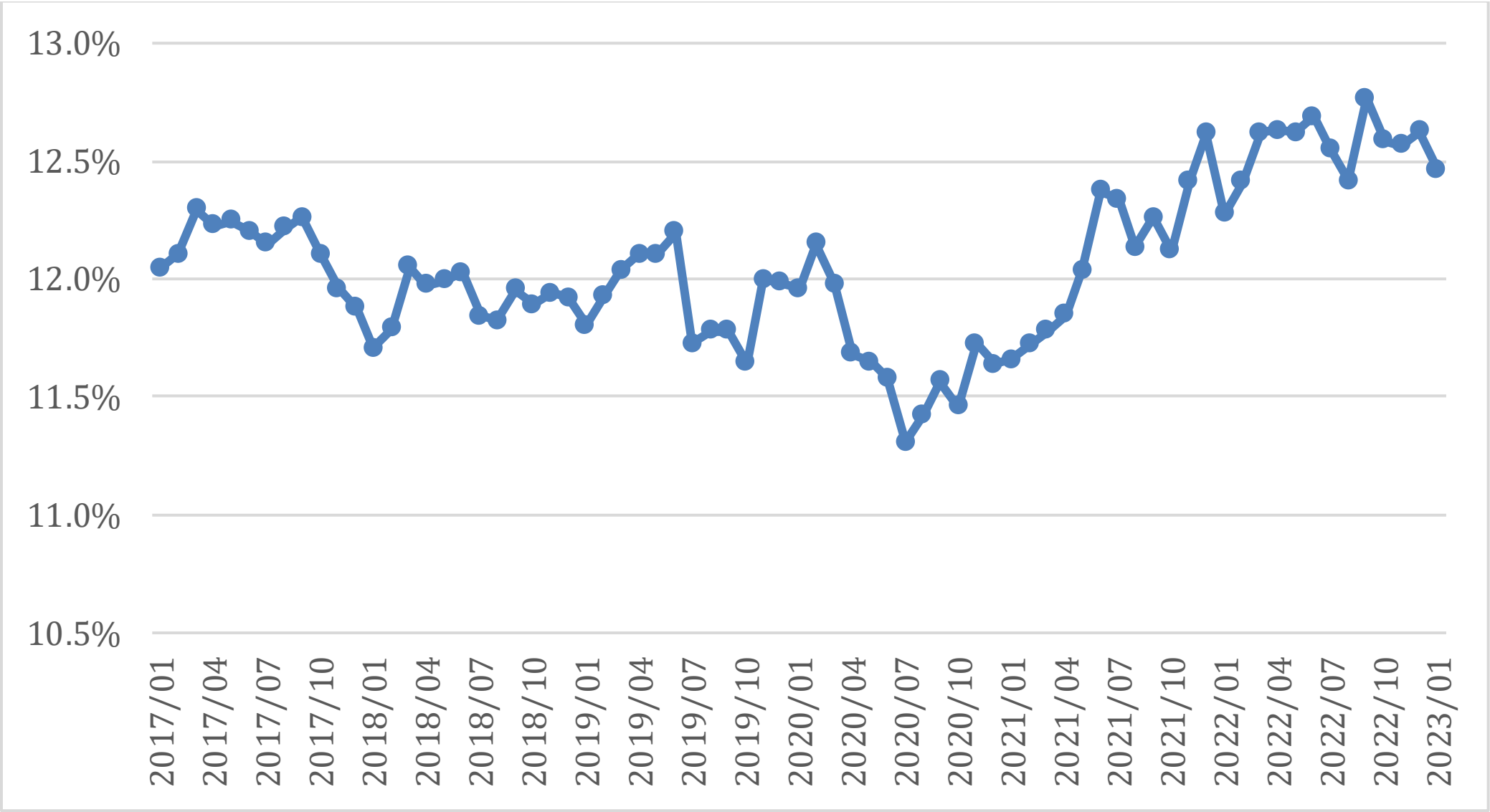

Constructed by the authors using monthly CPS data for all adults ages 18 and above, January 2017-January 2023.

Results in Figure 5 show that although women had consistently higher rates of disability throughout the pre-pandemic and pandemic periods, both men and women exhibited large deviations from the 2017-2019 trend during the pandemic period. Table 1 shows that, on average, disability rose by about 1 percentage point for both men and women in January 2022-January 2023 relative to the rate that would have been predicted by the 2017-19 trend. Most of that increase was accounted for by people reporting a cognitive impairment, although there were increases in other disability categories as well. While the gap between men and women's deviations in overall disability rates from trend is negligible, women have a larger deviation from trend compared to men for cognitive impairments. This category is most closely aligned to the brain fog symptoms commonly associated with long COVID, and the gender disparity is consistent with HPS data on both cognitive disability and long COVID.[7]

**Figure 5.** Deviations in Disability Rates from the Pre-Pandemic Trend by Gender

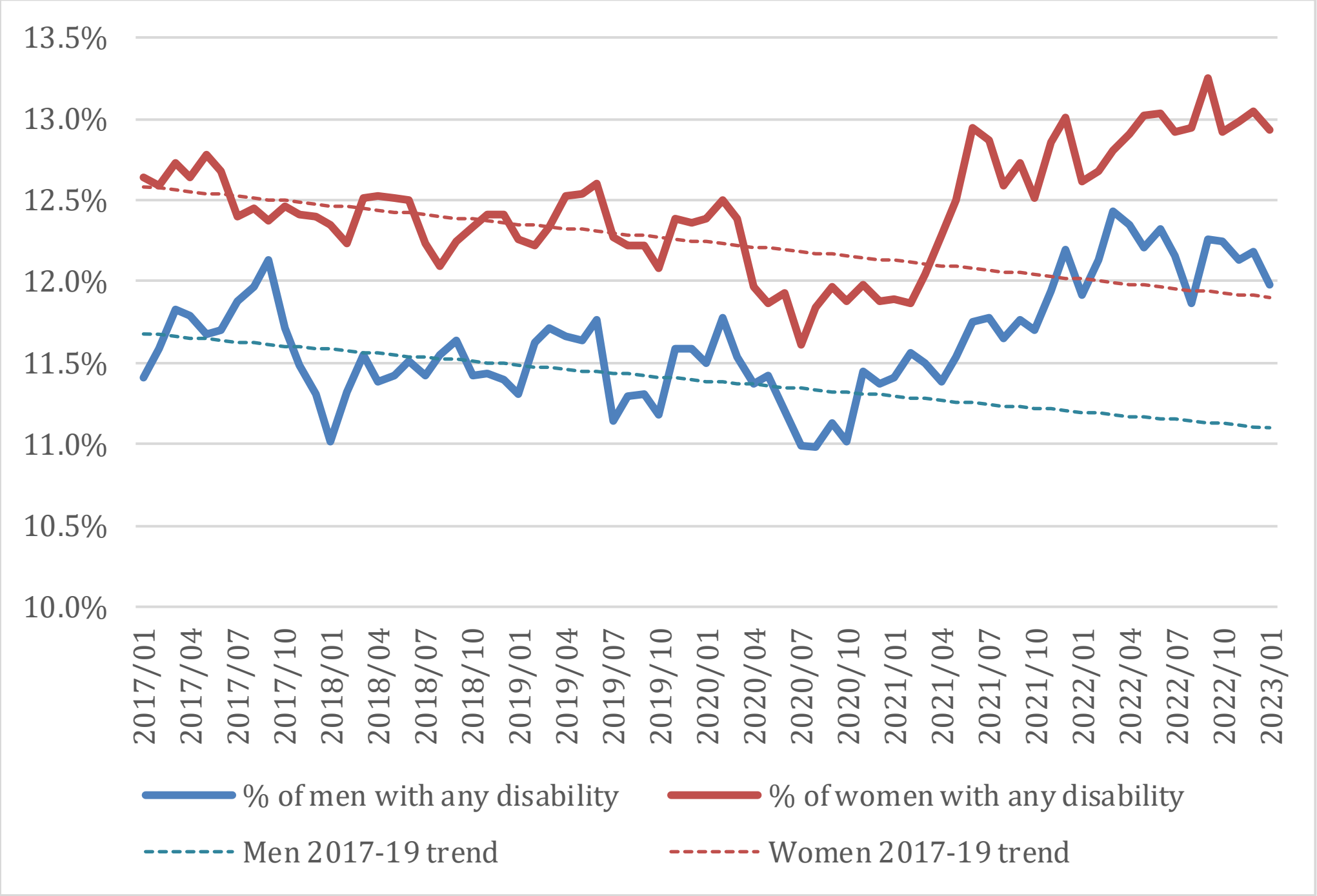

Constructed by the authors using monthly CPS data for all adults ages 18 and above, January 2017-January 2023.

**Table 1.** Change in Disability Rates among U.S. Adults Jan. 2022-Jan. 2023 Relative to 2017-2019 Trend (in percentage points)

| | *Total* | *Hearing* | *Visual* | *Cognitive* | *Mobility* | *Self-Care* | *Going Outside* |
|---|---|---|---|---|---|---|---|
| Men | 1.02 | 0.45 | 0.18 | 0.34 | 0.32 | 0.08 | -0.04 |

[7] Across the ten rounds of HPS data we analyzed, 6.5% of women reported severe difficulty remembering or concentrating compared to 4.5% of men.

| | | | | | | | |
|---|---|---|---|---|---|---|---|
| Women | 0.97 | 0.24 | 0.29 | 0.53 | 0.11 | 0.05 | 0.24 |
| | | | | | | | |
| White | 0.86 | 0.36 | 0.14 | 0.42 | 0.06 | 0.08 | 0.11 |
| Non-white | 1.45 | 0.25 | 0.56 | 0.50 | 0.71 | -0.01 | 0.06 |
| | | | | | | | |
| Younger (<45) | 0.81 | 0.14 | 0.19 | 0.60 | 0.16 | 0.09 | 0.15 |
| Older (45+) | 1.11 | 0.50 | 0.27 | 0.30 | 0.22 | 0.04 | 0.05 |
| | | | | | | | |
| Less educated | 0.93 | 0.33 | 0.24 | 0.44 | 0.18 | 0.04 | 0.11 |
| More educated | 0.98 | 0.34 | 0.20 | 0.38 | 0.17 | 0.10 | 0.05 |

Constructed by the authors using monthly CPS data for all adults ages 18 and above, January 2017-January 2023. Less educated is less than a four-year college degree, more educated is a college degree or higher.

We applied the same procedure to deviations in disability rates from trend for several other demographic groups with similar results regarding the fairly large increase in cognitive disabilities. Non-white individuals had a much higher increase in the overall disability rate compared to white individuals, with mobility, visual, and cognitive disabilities accounting for most of that increase. Of note, younger people had a relatively larger increase in cognitive disabilities, while older people experienced a larger increase in hearing disabilities. This latter finding is supported in the medical literature with evidence showing that because the inner ear is vulnerable to viruses, long COVID is associated with sensorineural hearing loss, tinnitus, and vertigo [19]. Similar to the HPS results, people without college degrees experienced a larger increase in cognitive impairments compared to people with more educational attainment.

We next turn to our regression analysis of remote work using the CPS, and to put that analysis into context, Panel A of Figure 6 shows that for much of the pandemic period, up through the end of 2021, the likelihood of working remotely because of the pandemic was lower for people with disabilities compared to people without disabilities. For example, in July 2021, COVID-19 caused 11.3% of people with disabilities to have to work from home compared to 13.5% of people without disabilities. This is in contrast to the period before the pandemic started, when people with disabilities had a greater likelihood of working from home as their usual work mode, as documented in Ameri et al. (2022) [20] and shown in Figure 6, Panel B. Until 2019, a higher percentage of people with disabilities worked remotely as their regular mode of employment compared to people without disabilities. The trend was reversed with the onset of the pandemic when lockdowns and social distancing protocols forced many employers to permit work-from-home options for their workers if the work could be performed remotely. This sudden common acceptance of remote work helps to explain why the proportion of people without disabilities who work remotely increased dramatically when the pandemic started. As shown in Figure 6 Panel A, the gap was as large as 10 percentage points in early 2020. Regression results in Column 1 of Table 2 support this explanation. When controlling only for the survey month, on average, people with disabilities were 2.3 percentage points less likely than people without disabilities to engage in pandemic-related remote work.

**Figure 6.** Percent of Workers Engaged in Remote Work

Panel A. Percent of Workings Engaged in Remote Work Specifically due to Pandemic, by Disability Status, May 2020-Sept. 2022 (Monthly CPS Data).

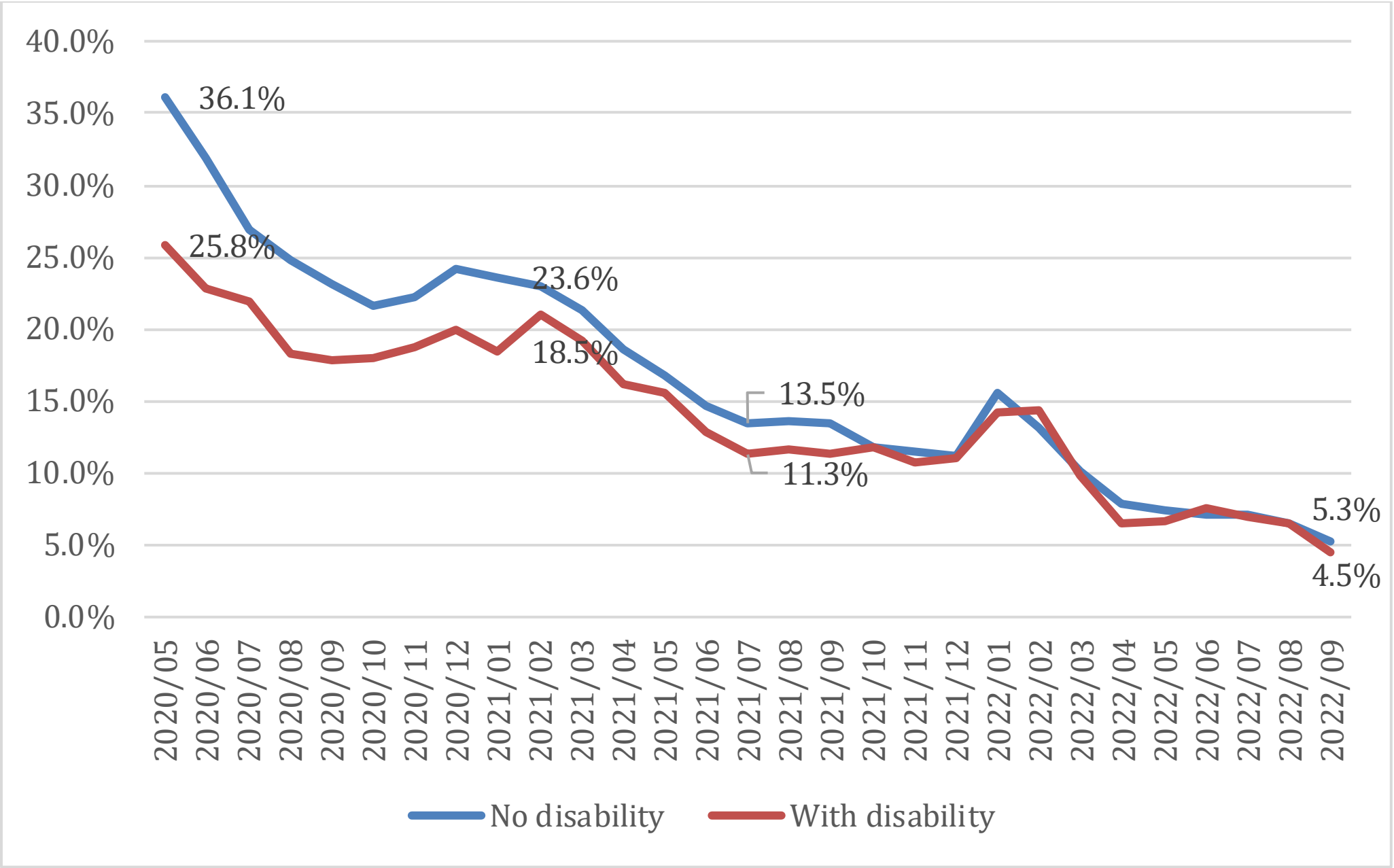

Constructed by the authors using monthly CPS data for all working adults ages 18 and above.

Panel B. Percent of Workers Working Primarily at Home Pre-Pandemic, by Disability Status, 2008-2020 (Annual data from American Community Survey)

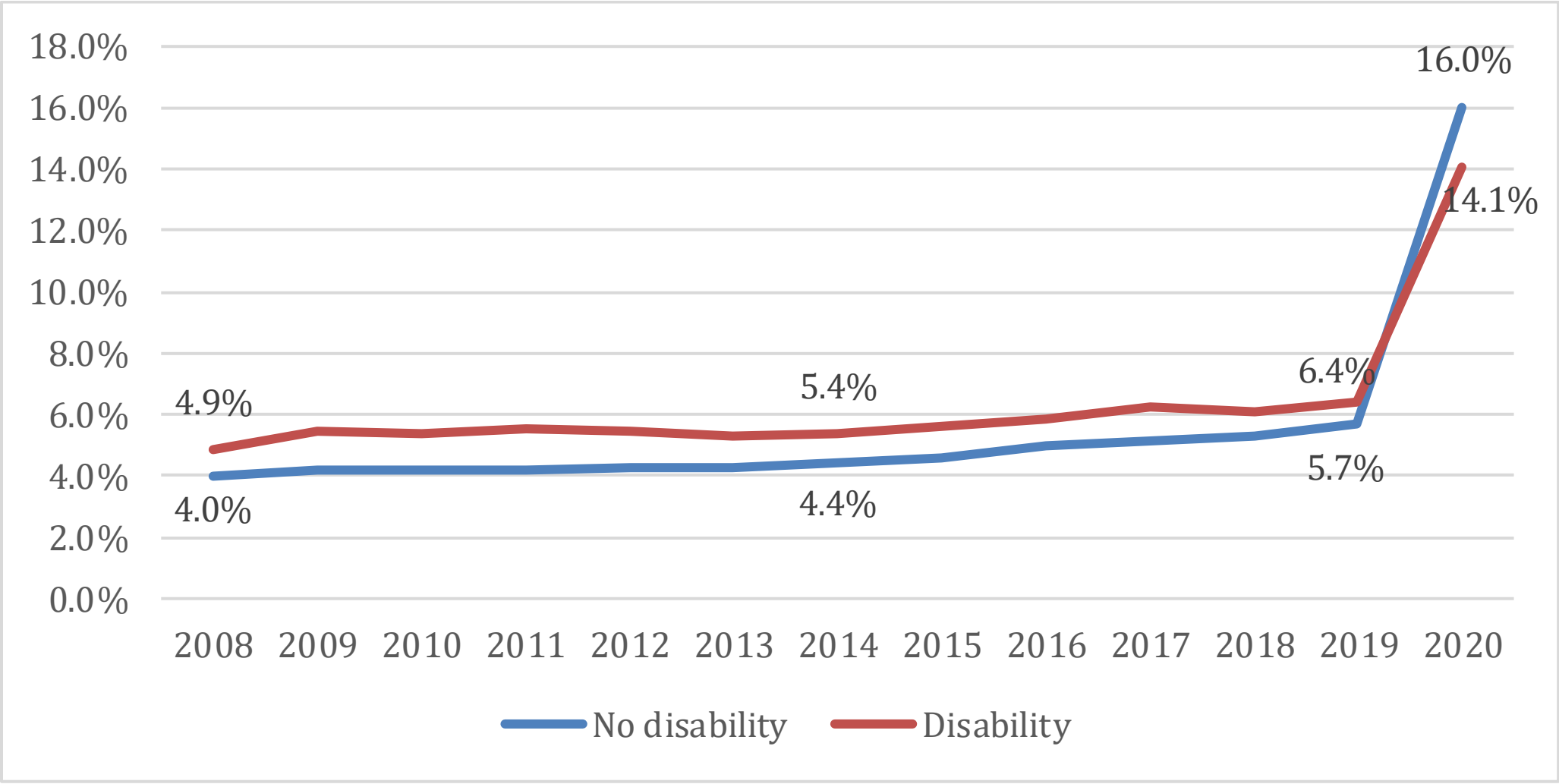

Replicated from Ameri et al. (2022) [20].

**Table 2.** Likelihood of Pandemic-Related Remote Work (Using Monthly CPS data, May 2020-September 2022)

| | *(1)* *Month Only* | *(2)* *Month + Occ/Ind* | *(3)* *Month + Dem. Char.* | *(4)* *Month + Occ/Ind + Dem. Char.* | *(5)* *Sep. Disability + Full Controls* |
|---|---|---|---|---|---|
| Disability | -0.023** | 0.000 | 0.016** | 0.016** | |
| | (0.003) | (0.002) | (0.003) | (0.002) | |
| Disability type | | | | | |
| Hearing impairment | | | | | 0.003 |
| | | | | | (0.002) |
| Visual impairment | | | | | -0.002 |
| | | | | | (0.004) |
| Cognitive impairment | | | | | 0.031** |
| | | | | | (0.003) |
| Mobility impairment | | | | | 0.013** |
| | | | | | (0.003) |
| Difficulty dressing or bathing | | | | | -0.008 |
| | | | | | (0.007) |
| Difficulty going outside alone | | | | | 0.006 |
| | | | | | (0.005) |
| Woman | | | 0.019** | 0.018** | 0.018** |
| | | | (0.001) | (0.001) | (0.001) |
| Race/ethnicity (white non-Hispanic excluded) | | | | | |
| Black non-Hispanic | | | -0.013** | 0.003** | 0.003** |
| | | | (0.002) | (0.001) | (0.001) |
| Hispanic/Latinx | | | -0.014** | 0.002** | 0.002** |
| | | | (0.002) | (0.001) | (0.001) |
| Other race/ethnicity | | | 0.059** | 0.045** | 0.045** |
| | | | (0.003) | (0.001) | (0.001) |
| Age (18-34 excluded) | | | | | |
| Age 35-49 dummy | | | 0.010** | 0.000 | 0.000 |
| | | | (0.002) | (0.001) | (0.001) |

| | (1) | (2) | (3) | (4) | (5) |
|---|---|---|---|---|---|
| Age 49-64 dummy | | | -0.005** | -0.015** | -0.015** |
| | | | (0.002) | (0.001) | (0.001) |
| Age 64-99 dummy | | | -0.022** | -0.032** | -0.031** |
| | | | (0.002) | (0.001) | (0.001) |
| Education (no HS degree excluded) | | | | | |
| High school degree/GED | | | 0.021** | -0.004** | -0.004** |
| | | | (0.001) | (0.001) | (0.001) |
| Associate's degree or some college | | | 0.069** | 0.010** | 0.010** |
| | | | (0.001) | (0.001) | (0.001) |
| Bachelor's degree | | | 0.220** | 0.081** | 0.081** |
| | | | (0.002) | (0.001) | (0.001) |
| Graduate degree | | | 0.303** | 0.145** | 0.145** |
| | | | (0.002) | (0.002) | (0.002) |
| Number of children under age 18 | | | -0.004** | -0.002** | -0.002** |
| | | | (0.001) | (0.000) | (0.000) |
| Part-time worker | | | -0.062** | -0.029** | -0.029** |
| | | | (0.001) | (0.001) | (0.001) |
| Self-employed | | | -0.039** | -0.040** | -0.040** |
| | | | (0.002) | (0.001) | (0.001) |
| Constant | 0.358** | 0.273** | 0.230** | 0.248** | 0.248** |
| | (0.003) | (0.005) | (0.003) | (0.005) | (0.005) |
| Month/year dummies | Yes | Yes | Yes | Yes | Yes |
| 524 occupation dummies | No | Yes | No | Yes | Yes |
| 51 industry dummies | No | Yes | No | Yes | Yes |
| Adjusted R-squared | 0.043 | 0.232 | 0.149 | 0.248 | 0.248 |

Note: Authors' computations using CPS data for May 2020-September 2022 (the period when the CPS included the remote work question in their survey). N=1,381,695 in all regressions. The dependent variable is whether or not someone is working remotely. Based on linear probability regressions. Robust standard errors in parentheses. The notation ** is p<0.01, * is p<0.05.

The relatively lower likelihood of people with disabilities working remotely during the pandemic may be explained by the occupations and industries where they work, both in how amenable these jobs are to remote work and how vulnerable workers were to layoffs when the pandemic started. Results in Column 2 of Table 2 support this assertion. When we include detailed occupation and industry categories, the negative coefficient on the disability indicator changes to a non-significant coefficient that is effectively zero. We conclude that occupation and industry account for most of the disability gap in pandemic-related remote work. In other words, the lower overall rate of pandemic-related remote work among people with disabilities appears to be primarily due to their underrepresentation in knowledge- and white-collar jobs (which are more amenable to remote work and were subject to fewer cuts) and overrepresentation in service sector and blue-collar jobs (which are less amenable to remote work and were subject to more cuts) [21, 22].

Yet occupations and industries are also highly segregated by gender, race, and educational attainment, so the occupation and industry controls could be picking up some effects that are due to these dimensions of social stratification. In other words, demographic characteristics may help to explain why people without disabilities suddenly had a greater propensity to work remotely due to the pandemic compared to people with disabilities. When we include gender, race, education, age, and job security in the regression model without occupation and industry controls (Table 2 Column 3), the coefficient on disability status remains small (1.6 percentage points) but becomes positive and statistically significant. This slightly higher likelihood of engaging in pandemic-related remote work for people with disabilities in the third model is consistent with pre-pandemic patterns showing the relatively higher likelihood of remote work as the usual mode of work among people with disabilities (Figure 6 Panel B).

Column 3 shows that being a woman and being highly educated are positively associated with the likelihood of remote work due to the pandemic. In reconciling this relatively higher likelihood for women to work remotely with their higher likelihood of having long COVID, we argue that a likely explanation is their caregiving responsibilities and the associated risks of exposure to the virus. The ‘other race/ethnicity’ category – which includes Asian, multiracial, and other groups – also has a positive association with pandemic-related remote work. In contrast, being older, Black or Hispanic, self-employed, or a part-time worker is negatively associated with the likelihood of engaging in pandemic-related remote work. These results suggest that job security, higher education, and youth have the added benefit of protecting some workers by allowing them to work from home because of the pandemic, while Black and Hispanic workers were more likely to be employed in ‘essential’ jobs with higher risk of exposure to the virus. The fully specified model in Column 4 includes the demographic characteristics plus the occupation and industry controls, and here the disability effect stays the same at 1.6 percentage points. Some of the coefficient estimates for the demographic groups do change, especially for race/ethnicity and education, confirming that the occupation and industry controls are picking up some effects of social stratification.

Differences across disability types are analyzed in Column 5, which includes the complete set of demographic characteristics along with occupation and industry. In this model, we see that the higher likelihood of pandemic-related remote work for people with disabilities is explained mostly by people with cognitive impairments and mobility impairments. The coefficient on cognitive impairment, the category that most closely resembles the brain fog symptoms associated with long COVID, is particularly large. These results suggest that holding all else equal, during the pandemic employers were more willing or were compelled to allow remote work arrangements for workers with cognitive and mobility impairments relative to people without reported disabilities. However, they did not do this for people with other types of disabilities.

## V. Remote Work: An Unintentional Accommodation?

The regression results could reflect a change in employer practices related to disability. However, because the analysis does not allow us to observe employer policies in any direct way, the results could be an artifact of pandemic conditions and the refusal of workers to return to the physical office. A related possibility is that some people already had unaccommodated cognitive and mobility impairments, or they developed cognitive and mobility impairments from long COVID while remote work arrangements were in place because of the lockdowns. For the shift to reflect changes in disability policy, people would have to request and receive accommodations. In contrast, people who developed impairments during the pandemic period or already had impairments but never had accommodations may not have needed to request accommodations. When they allowed remote work during the pandemic, employers may not even have been aware that they were providing accommodations for employees with a disability; this may be a case of *unintentional accommodation*. This possibility is supported by pre-pandemic evidence indicating that if anything, accommodation requests from workers with mental health impairments were less likely to be seen as reasonable and granted by employers compared to requests from people with physical disabilities [23].

Unintentional accommodation via remote work could mark a sea change for people with mental health challenges like depression, anxiety, post-traumatic stress disorder, neurological and developmental disorders, and neurodiversity more broadly (including Tourette Syndrome, attention deficit disorder, autism spectrum disorder, dyslexia, dyspraxia, and sensory needs).

Reluctance to disclose such impairments is widespread [24]. The pool of people with a disability in high school declines in college and drops further in the workplace – not because people with disabilities do not go to college or work, but because they cease to identify as having a disability or choose not to report it. Nationally representative survey data indicate that almost 40% of high school students reported ever having a disability in 2016 [25].[8] Of those students with disabilities, 60% went to a 2- or 4-year college, but only 13% reported a disability to the college. In other words, 9,200 high school students had a disability, 5,520 of them went to college, but only 718 students reported a disability to the college. This figure is likely even smaller once people enter the labor force and navigate issues around disclosure with an employer. According to CPS-based estimates in Schur et al. (2023) [26], the average percentage of employees who self-identify as having a disability is 3.2%. Upwards of 80% or more employed adults with a disability have never informed their employer, much less sought accommodation, and roughly half of those adults could have a learning disability.[9]

These workers are overrepresented in blue-collar jobs such as building cleaning and transportation and in pink-collar jobs such as community service and healthcare support – jobs that are not conducive to home-based work [27]. For those white-collar workers who were able to work remotely because of the pandemic, remote work may be the first time since high school (if ever) they experienced an accommodation.

According to data from a nationally representative sample, among white collar employees, younger generations are more likely than older generations to have a disability – but less likely to disclose the disability to human resources [28]. Due to long COVID as well as an aging population, the proportion of people in the labor force with a disability is likely to continue to grow. Long COVID should kickstart a concerted planning effort in human resource management; one in which the onus shifts from people with disabilities to request accommodation to one in which employers include disability accommodations at all steps of the employment process. When they are able to disclose their disabilities, workers with disabilities are twice as likely to be happy or content with their jobs (65% versus 27%), half as likely to experience anxiety (18% versus 40%), and almost five times less likely to feel isolated (8% versus 37%) [28]. Employers should assume that large portions of their workforce, likely above 20%, have a disability and will be more productive and more satisfied with their jobs when accommodations are available.

[8] The 23,000 students were in 9th grade in 2009 and were surveyed in 2016; the data aggregate student self-reports, school reports on students with Individualized Education Programs, and parent reports of diagnoses by health or education professional [25].
[9] Calculation based on data in Newman et al. (2011) [25].

## VI. Accommodating People with Long COVID Impairments

COVID-19 and COVID-related impairments are not going away anytime soon, which means that employers need to thoughtfully and intentionally adapt their policies and practices to the increased prevalence of disability in the labor force. Although remote work is a major accommodation for people with activity limitations related to long COVID, it cannot be the only one given that people with disabilities are disproportionately employed in industries and occupations where remote work is not feasible [26]. In addition to remote work, accommodations such as shorter workdays, flexible scheduling, reduced workloads, less contact with customers/clients, deferment of more complex job responsibilities, task-swapping, shared responsibility for sets of tasks, and more rest breaks for people with long COVID-related impairments are critical. Changes to the physical workspace may also be warranted, including more ergonomic tools, specialized seats and carts, electronic organizers, and alternative lighting [29]. Additional workplace policies include having more COVID-sensitive absence policies, training managers and supervisors to be more knowledgeable and compassionate about long COVID, developing practices to offer accommodations for symptoms like fatigue that cannot be measured objectively, and promoting disability-friendly culture shifts [29, 30]. The unpredictability and intensity of some symptoms make such accommodations and changes in policies and practices particularly important. Failure to accommodate people can lead to injuries in the workplace, higher turnover, loss of firm-specific human capital, innovation shortfalls, and reduced productivity [27, 28].

Not all people with disabling chronic conditions related to COVID-19 may identify as having a disability, but people with long COVID should be aware of accommodations that address their activity limitations [31]. In the U.S., workplace accommodations are legally mandated in the context of the federal government's 1973 Rehabilitation Act, the first national law establishing civil rights protections for people with disabilities, and the 1990 ADA. Both set standards on how and when employers must provide workplace accommodations. However, courts have generally held that employers are not obligated to adopt worker requests for remote work and flexible scheduling, and many employers have remained resistant [32, 33]. Such resistance has serious consequences in the case of impairments related to long COVID. From a global perspective, employer resistance to providing accommodations may be less relevant in other countries such as the U.K. where workplace accommodations for disability are a more established practice or clearly advocated within national guidance [34].

Workers in low-wage jobs, including many of those impaired by long COVID, may find it particularly difficult to access or request accommodations, yet they are precisely those who need accommodating jobs the most to safeguard their economic security. People of color with long COVID face additional challenges from institutionalized racism in obtaining accommodations at work, with long-term consequences as persistent discrimination is associated with persistent disparities in health [2].

Results from this study show that in the U.S., women, Hispanic individuals, sexual and gender minorities, people with preexisting disabilities, and people with less education are more likely than their counterparts (men, non-Hispanic, cisgender and straight, people without disabilities, and highly educated individuals) to have long COVID and to report more activity limitations due to long COVID. Our analysis of social categories like gender, ethnicity, and sexuality suggests that disparities in COVID-related impairments have less to do with the virus itself and more to do with labor market stratification [35]. Not only did women have more activity limitations due to long COVID, they were also disproportionately represented in frontline occupations during the pandemic that were more exposed to contagion [36, 37]. Historically women, racial/ethnic minorities, sexual and gender minorities, and people with disabilities have been excluded from the full range of occupations in the U.S. and are concentrated in a smaller set of sectors. Further, given the overrepresentation of men and highly educated individuals in management positions (people who are least likely to have long COVID or to experience activity limitations from long COVID), the more severe impacts on other populations may seem 'unrelatable,' and those managers may be hesitant to make accommodations.

These results for long-Covid prevalence and associated activity limitations support arguments for a stronger social safety net (including paid family leave, sick leave, and disability benefits), especially because people with disabilities are more likely to hold low-wage, part-time, contingent, and gig jobs which generally do not provide paid leave and other benefits that people may enjoy in more secure jobs [38]. Jobs without an adequate social safety net place people with disabilities at higher risk for loss of employment, wages, independence, and economic self-sufficiency. The results also reveal systematic disparities in health over the long term. This is an urgent issue presenting ongoing challenges to already-disadvantaged people – exactly the kind of challenges that require policy mediation that the Household Pulse Survey was designed to inform – with implications for individuals, families, and communities. The occurrence of COVID-19, a mass disabling event, should sensitize everyone to the needs of people with disabilities. A myriad of people have undisclosed disabilities, and long COVID could tip the scale. The stigma and discrimination associated with disability that discourage many people from seeking accommodation could be mitigated if offering, requesting, and granting accommodations become part of workplace culture.

**Statements and Declarations**

Funding: Author YR has received funding support from the National Institute on Disability, Independent Living, and Rehabilitation Research (NIDILRR) for the Rehabilitation Research & Training Center (RRTC) on Employment Policy: Center for Disability-Inclusive Employment Policy Research Grant [grant number #90RTEM0006-01–00] and by the NIDILRR RRTC on Employer Practices Leading to Successful Employment Outcomes Among People with Disabilities Research Grant [grant number #90RTEM0008-01-00].

Competing Interests: The authors have no relevant financial or non-financial interests to disclose.

Data Availability: The datasets generated during and/or analyzed during the current study are available from the corresponding author on reasonable request.

Ethical Statement: This study was performed in line with the principles of the Declaration of Helsinki. Approval was granted by the Institutional Review Board of Rutgers University, October 20, 2022, Study ID Pro2021002068.